\documentstyle[preprint,tighten,aps]{revtex}

\begin{document}

\def\Jo#1#2#3#4{{#1} {\bf #2}, #3 (#4)}
\def \re#1{(\ref{#1})}
\def\st{\scriptstyle}
\def\sst{\scriptscriptstyle}
\def\mco{\multicolumn}
\def\epp{\epsilon^{\prime}}
\def\vep{\varepsilon}
\def\ra{\rightarrow}
\def\vp{{\bf p}}
\def\al{\alpha}
\def\ab{\bar{\alpha}}
\def \bi{\bibitem}
\def \ep{\epsilon}
\def\D{\Delta}
\def\sms{$\s$-models }
\def \om {\omega}

\def\be{\begin{equation}}
\def\ee{\end{equation}}
\def \lab {\label}
\def \k {\kappa} 
\def \del {\partial}
\def \bd {\bar \partial }
\def \na {\nabla}
\def \const {{\rm const}}
\def \ha{{\textstyle{1\over 2}}}
\def \na {\nabla }
\def \D {\Delta}
\def \a {\alpha}
\def \b {\beta}
\def \chi {\chi}\def\r {\rho}
\def \s {\sigma}
\def \p {\phi}
\def \m {\mu}
\def \n {\nu}
\def \vp {\varphi }
\def \l {\lambda}
\def \t {\theta}
\def \td {\tilde }
\def \d {\delta}
\def \ci {\cite}
\def \la {\label}
\def \sm {$\s$-model }
\def \foot {\footnote }
\def \P {\Phi}
\def \o {\omega}
\def \inv {^{-1}}
\def \ov {\over }
\def \four{{\textstyle{1\over 4}}}
\def \fourth{{{1\over 4}}}
\def \foot{\footnote}
\def\be{\begin{equation}}
\def\ee{\end{equation}}
\def\bea{\begin{eqnarray}}
\def\eea{\end{eqnarray}}
\def\np {{\em  Nucl. Phys. }}
\def \pl {{\em  Phys. Lett. }}
\def \mpl {{\em Mod. Phys. Lett. }}
\def \prl {{ \em  Phys. Rev. Lett. }}
\def \pr  {{\em  Phys. Rev. }}
\def \ap  {{\em Ann. Phys. }}
\def \cmp {{\em Commun.Math.Phys. }}
\def \ijmp {{\em Int. J. Mod. Phys. }}
\def \jmp {{\em J. Math. Phys.}}
\def \cqg {{\em  Class. Quant. Grav. }}

%%%%%%%%%%%%%%%%%%%%%%%%%%%%%%%%
%%%%%%%%%%%%%%%%%%%%%%%%%%%%%%%%%%
\preprint{Imperial/TP/95-96/65, \  gr-qc/9608044} 
\date{August 1996}
\title{COMPOSITE  BLACK HOLES IN STRING THEORY\thanks{Talk at the Second International Sakharov Conference on Physics, Moscow,  May 20-24, 1996.}
}
\author{ A.A. TSEYTLIN }
\address{Blackett Laboratory, Imperial College, London SW7 2BZ, U.K.\\
and\\
Lebedev Physics Institute, Moscow,  Russia }
%%%%%%%%%%%%%%%%%%%%%%%%%%%%%%%%%%%%%%%%%%%%%%%%%%%%%%%%%%%%%%
\maketitle
\begin{abstract}
{We  discuss  properties  of special supersymmetric 
extreme black holes  in 4 and 5 dimensions  which have regular 
horizons, non-zero entropy and can be interpreted as 
compactifications of  BPS bound states of p-branes in 10 or 11 dimensions. } 
% some recent progress  in .... a universal relation between 
% the black hole  entropy (area) and  the 
%statistical entropy of BPS-saturated oscillation states 
%of solitonic string.}
\end{abstract} 
%\end{titlepage}

%\newpage 
%%%%%%%%%%%%%%%%%%%%%%%%%%%%%%%%%%
%\section{Introduction}
%%%%%%%%%%%%%%%%%%%%%%%%%%%%%%%%%%
\bigskip\bigskip

Recently,  black holes in string theory  have become a subject of
intensive research.  It was realised that  microscopic
properties (e.g.  statistical origin of the   entropy)
of certain  composite supersymmetric black holes  can be addressed 
systematically  using  either conformal field theory description  of 
 the NS-NS backgrounds or $D$-brane  representation  of U-dual  R-R backgrounds
(for reviews see \cite{TMpl,Horowitz}).

Low-energy effective actions of string theories 
contain the metric and a collection of vector and scalar fields.
The corresponding equations have various black hole solutions
with several vector and scalar fields being non-trivial.
Solutions without residual supersymmetry 
(like Schwarzschild one)  are, in general,  
 deformed by $\a'$-corrections, so that  their  
properties are  hard to determine exactly.  
Most of supersymmetric  backgrounds 
 have certain types of singularities at the horizon.
For example, the metric of  purely electric extreme  black hole  has  
 horizon coinciding with singularity (suggesting that $\a'$ corrections should become important there),  while purely magnetic 
extreme black hole has dilaton scalar blowing up at the horizon
(so that one cannot  a priori ignore  string loop corrections).
All extreme black holes with singular horizons have vanishing 
area of the horizon, i.e.
zero semiclassical Bekenstein-Hawking (BH) entropy. 

There are, however,  remarkable exceptions: 
special extreme (BPS saturated)  black holes 
with  four independent charges in $D=4$ and three independent charges in $D=5$
have regular horizons
 and non-zero BH entropy \cite{CY,kall}.
They have minimal possible amount ($N=1$) of 
residual supersymmetry.
  The fact that all scalars  are regular and  approximately constant
makes them  look `realistic'. 
 They are thus  closest analogues of the standard 
extreme  Reissner-Nordstr\"om solution of Einstein-Maxwell theory
(for equal values of charges they, indeed, can be viewed as embeddings 
of RN solution into  string theory). 
This opens a possibility of understanding  of their 
properties directly from string theory.

These solitonic solutions have several surprising  and exceptional features.

(1) They exist only in dimensions $D=4$  and $D=5$  supported by  at least 
$n=4$ and $n=3$ different vector fields  with  the associated 
charges $Q_i$, $ i=1,..., n$  (some of which may be  electric, and some  - magnetic).
The corresponding $D$-dimensional Einstein-frame metric is
\begin{equation}
ds^2_D=-\lambda^{D-3}(r) dt^2+\lambda^{-1}(r)\big[dr^2+r^2d\Omega_{D-2}
^2\big]\  ,
\label{one}\end{equation}
\begin{equation}
\lambda (r)= (H_1...H_n)^{-{1\over {D-2}}}\ , 
\ \ \ \ \ 
H_i = 1 + { Q_i \ov r^{D-3} } \ . 
\label{two}
\end{equation} 
The one-center solution with all charges at $r=0$ has straightforward 
extension to the  case of more general   harmonic functions
$H_i$, e.g.,  with  centers at different points.
 The RN case corresponds to $Q_i=Q$.
The ADM mass is 
\be
 M  = b (Q_1 + ... + Q_n) \ , 
\ \ \ \    b \equiv 
{\omega_{D-2}\ov 2\k^2_D } (D-3)  \ , 
\la{mas}
\ee
where $\k_D$ is gravitational constant in $D$ dimensions
and $\omega_{D-2}$ is $4\pi$ in $D=4$ and $2\pi^2$ in $D=5$.
The fact that these solutions saturate the BPS bound is illustrated
by the expression for the mass of their  non-extreme versions
($\m$ is a non-extremality parameter
 which is zero in the extreme limit and is 
 proportional to the 
Schwarzchild mass in the case of $Q_i=0$, i.e. 
it enters through the function $f=1 - {2\m\ov r^{D-3}}$): 
\be
 M  = b [ (Q_1^2  + \m^2)^{1/2}  +  ... + ( Q_n^2  + \m^2)^{1/2}  ]   \ . 
\la{mase}
\ee
The latter  is reminiscent of the energy of a system of particles 
with masses $Q_i$ boosted to the same momentum $\m$. 
The BPS-saturated configuration can be interpreted as a bound state with zero binding  energy. 
Indeed, the mass remains the same for  the multicenter solution,  suggesting  that 
charges can be separated at no cost in the energy. 
This `composite object'  interpretation  
is   explained  by embedding these solutions into $D=10$ or $11$ dimensional theory as discussed below \cite{TM,KTT}.

(2)  The  $(D-2)$-sphere  at $r=0$  is a regular horizon with finite 
area $A_{D-2}$; the corresponding BH entropy is 
\be
S_{BH}= {2\pi A_{D-2}\ov \k^2_D}
 = c \sqrt {Q_1 ... Q_n} 
\ ,  \ \ \ \ \ \  \ \ c\equiv {2\pi\omega_{D-2} \ov \k^2_D}  \ . 
\la{arrq}
\ee
Being proportional to the product of charges the entropy vanishes
unless  there is a 
 `critical' number of non-zero charges ($n=4$ for $D=4$ and $n=3$ for $D=5$).
The entropy vanishes for multicenter solution; 
the single-center case is a point of enhanced (spherical) symmetry
where the entropy becomes non-vanishing.
The non-extremal  version  has the entropy \ci{CVT} 
$$
S_{BH}=   c     [(Q_1^2 + \m^2 )^{1/2} +  \m]^{1/2}  ...  \  [  (Q_n^2 + \m^2 )^{1/2} +  \m]^{1/2}  \ . $$ 
The charges are expected to  take quantized 
values in the quantum theory, $Q_i \propto m_i$, where $m_i$ are integers
(the quantization condition is fixed by the  embedding into 
10-dimensional string theory or 11-dimensional M-theory).
Expressed in terms of {\it integers}  $m_i$ the  BH entropy \re{arrq}
takes simple and universal  form  \ci{CTII}
\be
S_{BH} = 2\pi  \sqrt {m_1 ... m_n} \ ,  
\la{uni}
\ee
suggesting a  possibility of a statistical interpretation (see, e.g., \cite{Sen,LW,SV,CM,MS,TLast}).
It can be indeed be interpreted (for large 
values of charges)
 as   the leading term in the  entropy $S = \ln d(m)$
associated with supersymmetric BPS states of  an effective 
$D=6$ supersymmetric solitonic string  with momentum number $m=m_1$
and winding number (along  compact 6-th direction)
$w= m_2...m_n$. The entropy
of  $N= N_B + \ha N_F $  massless particles moving  in one direction 
in 1+1  dimensions
(with  compact space of large length $L$) 
is $S= \sqrt {{\pi\ov 3}  N E L}$, $E= 2\pi mw/L$. 
For a string in $D=6$  one has $N_B=N_F=(D-2)=4$ (i.e. $c_{eff} =6$). 
One puzzling feature of \re{uni}
is the rapid growth of the entropy with charges.
This may be related to the string-soliton nature of the solution.
The standard euclidean path integral derivation of the 
BH entropy starts with semiclassical field-theory partition 
function which itself may be viewed as an exponent 
of the first-quantized string partition function
(note that similar exponentiation is needed
to combine the string source action with space-time action
in order to discuss  solutions supported by fundamental string sources).
This may be suggesting  to represent
$S_{BH}$ as $ 2\pi exp [  \ha  (\ln m_1 + ... + \ln m_n)]$
and to interpret  $\ln m_1$ as `partial' statistical entropies.

(3)  These $D=4,5$  black hole backgrounds  admit   several 
possible embeddings into  10-dimensional  string theory
and 11-dimensional supergravity (or M-theory).

(3a) If all the charges  $Q_i$ are of NS-NS nature 
the corresponding bosonic background 
(which is a solution of either heterotic or type II superstring theory)
 is represented 
by  the direct sum of supersymmetric 
6-dimensional conformal sigma-model
and   free 4-torus  model  \ci{CTII}.
It has the form of a  `superposition' 
of the fundamental string model (with non-trivial time-like part
incorporating two electric charges, related to winding and momentum 
of the fundamental string source)
and the 5-brane-type model (with non-trivial (4,4) supersymmetric
`hyperk\"ahler space with torsion' transverse part 
depending on one or two 
magnetic charges  for $D=5$ or $4$ black hole).
%magnetic charge regularises short-distance behaviour
For example, the  $D=10$ conformal model associated 
with $D=5$ black hole is
$$   L =  F(x)  \del u \left[\bd v +
   K(x) \bd u \right] + f(x)  \del x^m \bd x^m 
$$
\be
 +  \     B_{mn} (x) \del x^m \bd x^n + 
    \del y^n \bd y^n  +  {\cal R} \P(x) \ ,   
\la{opi}
\ee
where $m,n=1, ...,4$,\  $u=y_5-t, v=y_5 +t$, 
\ $
 H_{mnk} = - \epsilon_{mnkl} \del_l  f ,
$ $e^{2\P}= fF$, and $F\inv, K$ and $f$ are harmonic functions, 
$F\inv =1 + Q_1/r^2$, etc.  The  regular $D=5$ black 
hole with 3 NS-NS 
charges can thus be interpreted as a dimensional reduction 
of a  BPS bound state of a fundamental string and solitonic
 5-brane, with 5-brane  wrapped around  5-torus
and the string wound around one circle of the torus
with extra momentum flow along the string ($Q_1,Q_2,Q_3$ are thus
proportional to the winding number, string momentum and 5-brane charge).
Dropping the trivial 4-torus part of 5-brane  one gets 
a  conformal model describing $D=6$ solitonic string.
The BH entropy can  then  be understood as the  
leading  universal term in the statistical  entropy of 
$D=6$ string  associated with supersymmetric marginal perturbations 
$ {\cal A}_m (x,u) \del u \bd x^m$ in 
4 non-compact directions $x^m$  (perturbations
in compact $y^n$ directions give only subleading 
contributions to the entropy).
These  more general deformed models describe a family
of supersymmetric black holes
which all have the same large-distance  form (same asymptotic charges)
but different short-distance structure.
Since perturbations decay at large distances
their number can be counted by considering the
 model near $r=0$ where it becomes equivalent to 
$SL(2,R) \times SU(2)$ WZW model with level equal to the magnetic 
charge $Q_3$. $Q_3$-factor effectively rescales 
the tension of the solitonic string as compared to the free fundamental string, making the associated entropy 
proportional to $\sqrt {Q_1Q_2Q_3}$, and thus matching the BH expression \ci{TLast}.

(3b)  The type II superstring background 
corresponding to the case when  (some of) the charges are of R-R type 
 can be obtained from the above NS-NS background 
 by  applying  $SL(2,R)$ and $T$   duality.
The simplest example is the description of $D=5$ black hole 
as a bound state of R-R string lying  within R-R 5-brane
of  type IIB theory. 
The corresponding  type IIB string-frame metric 
is
\be
ds^2_{10} = (fF\inv)^{1/2} [ F f\inv (dudv + Kdu^2) +  f\inv dy^n dy^n
+ dx^m dx^m ] \ .  
\la{oo}
\ee
The number of  associated excited BPS states 
and thus the statistical entropy can  then be evaluated 
by representing this R-R background in terms of $D$-branes \ci{SV,CM,MS}.

 There are various other U-dual representations
 in terms of configurations of
intersecting   NS-NS (solitonic) or R-R  p-branes in $D=10$. These
can be also described as anisotropic 6-branes (in $D=4$ black hole case)
or 5-branes (in $D=5$ black hole case)
with all internal coordinates being compact.
The most interesting example is  provided by the configuration of 
four  orthogonally 
intersecting 3-branes  ($3\bot 3\bot 3 \bot 3$) of type IIB 
theory, each one intersecting each of the three  others over a line.\foot{The number of intersecting objects cannot be greater 
than 4 if one demands at least $SO(3)$ isometry in non-compact directions.}
This remarkably symmetric 
configuration \ci{KTT,BL} can be pictured,  symbolically,  as a pyramid 
with 4 triangles standing for 3-branes, each pair sharing one  
dimension. The corresponding metric is 
$$
ds^2_{10}= (H_1H_2H_3H_4)^{1/2}  
\big[  - (H_1H_2H_3H_4)\inv  dt^2    + (H_1H_2)\inv dy_1^2  +   (H_1H_3)\inv dy_2^2
$$  $$ +    
 (H_1H_4)\inv dy_3^2 +   (H_2H_3)\inv dy_4^2+   (H_2H_4)\inv dy_5^2+ 
  (H_3H_4)\inv  dy_6^2+  
 dx^s dx^{s}\big] \ ,      $$
where $s=1,2,3$ and  $H_i = 1 + {Q_i\ov r}$ are harmonic functions
corresponding to each of four  3-branes. 
Dimensional reduction  along  6 compact internal directions
$y^a$ leads to  regular $D=4$ black hole background.
Scalar fields come from internal components of the metric (which are constant
if all charges are equal)
while the vector fields originate from  non-vanishing 
self-dual 5-tensor background \ci{KTT}.
An  interesting open problem  is how 
to compute the associated statistical entropy in 
way which is manifestly symmetric with respect to all 4 charges.

(3c) Another indication of the fundamental nature of the regular $D=4$
and $D=5$ black holes is that they have  a very  simple representation
\ci{TM,KTT}  in terms
of intersections of    basic  `M-branes' of 11-dimensional supergravity:
2-brane \ci{DS} and 5-brane \ci{Guven}.
The $D=5$ black hole  can be  obtained by dimensional 
reduction of  the symmetric $ 2\bot 2\bot 2$ configuration
(with all 2-branes intersecting at a point) 
or  of $2\bot 5$ one  (with 2-brane and 5-brane intersecting over a string with an extra momentum flow along it) \ci{TM}.  Similarly, 
the $D=4$ black hole  can be  obtained from 
compactified  anisotropic 7-brane represented   either by 
 $ 2\bot 2\bot 5\bot 5$ configuration
or  by    $5\bot 5\bot 5 $ configuration
 with  the fourth charge 
being related to the momentum along 
the string common to all three 5-branes.
The $D=11$ background for   $ 2\bot 2\bot 5\bot 5$  
is \ci{KTT} 
$$
d s^2_{11} =  (T_1T_2)^{-1/3} (F_1 F_2)^{-2/3}
 \big[ - T_1 T_2 F_1 F_2  \ dt^2  
+  T_1 F_1  dy_1^2 +  T_1 F_2  dy_2^2  $$
\be    
 + \ T_2 F_1  dy_3^2 +  T_2  F_2  dy_4^2
+  F_1F_2 ( dy_5^2 + dy_6^2 + dy_7^2) 
+  dx_s dx_{s}\big] \ ,     
\la{ooi}
\ee
$$ {\cal F}_4= -dt\wedge d (T_1dy_1\wedge d y_2 +T_2 dy_3\wedge d y_4 )  +
 *dF\inv_1 \wedge dy_2 \wedge dy_4 + *dF\inv_2 
\wedge dy_1 \wedge dy_3   . $$
Here $ T_i\inv = 1 + {Q_i\ov r} , $ and 
$  F_i\inv = 1 + {P_i \ov  r}$ are the inverse powers of harmonic functions 
associated with 2-branes and 5-branes, and ${\cal F}_4$ 
is the field strength of the antisymmetric 3-tensor field. 
The M-theory arguments leading to the expression \re{uni}
for the corresponding 7-brane entropy in terms of quantised charges
(in $D=11$ there is just one antisymmetric tensor and thus a unique quantisation condition) and its interpretation as a statistical entropy 
(for configurations $2\bot 5$ and $5\bot 5\bot 5$ with common string) 
were   discussed  in \ci{KTT}.
The  11-dimensional theory  understanding 
of the entropies of the corresponding near-extremal configurations 
was  presented  in  \ci{KKTT}.

To summarize, 
   extreme black holes  which have  regular horizons
in string theory  are 
 lower-dimensional images of composite extended objects 
 wrapped around compact internal dimensions
(e.g. solitonic string in  6 dimensions  or  special 
supersymmetric configurations of intersecting 
p-branes in 10 or 11 dimensions).
The existence of  a  family of  supersymmetric
 black holes  (all having  the same asymptotic charges
but different short-distance structure at  compactification scale) 
which correspond to `excited' or `oscillating'
states  of underlying extended objects 
provides a natural explanation for a non-vanishing BH entropy.
The fact that the expression for the universal
(large charge limit) part of the statistical entropy
can be understood  in terms of effective $D=6$ {\it string}
(which appears in  all --
NS-NS conformal sigma model,  R-R D-brane and M-brane -- approaches)
 has probably  more to do with supersymmetry 
(requirement of BPS   property) than with fundamental string theory.
This is supported  by the existence of a heuristic M-theory
explanation of the entropy (in terms of massless modes `living' 
on intersection  sub-spaces), which complements  the  NS-NS conformal sigma model  and $D$-brane  arguments.

\medskip

I am grateful to the organizers  of the  Sakharov's 
conference for their  excellent work 
and acknowledge also the support of PPARC and 
ECC grant SC1$^*$-CT92-0789.

%%%%%%%%%%%%%%%%%%%%%%%%%%%%%%%%%%%%%%%%%%%%%%%%%%%%%%%%%%%%
%%%%%%%%%%%%%%%%%%%%%%%%%%%%%%%%%%%%%%%%%%%%%%%%%%%%%%%%%%%%%%
% \section*{Acknowledgments}
%%%%%%%%%%%%%%%%%%%%%%%%%%%%%%%%%%%%%%%%%%%%%%%%%%%%%%%%%%%%

 %\section*{References}
%\begin{thebibliography}{99}

\end{document}